\pdfoutput=1

\documentclass[twoside,12pt]{article}
\usepackage{graphicx}

\def\lsim{\mathrel{\rlap{\lower3pt\hbox{\hskip1pt$\sim$}}
     \raise1pt\hbox{$<$}}} 
\def\gsim{\mathrel{\rlap{\lower3pt\hbox{\hskip1pt$\sim$}}
     \raise1pt\hbox{$>$}}} 

\newcommand{\be}{\begin{equation}}
\newcommand{\ee}{\end{equation}}
\newcommand{\ba}{\begin{eqnarray}}
\newcommand{\ea}{\end{eqnarray}}

\begin{document}





\title{Toward the AdS/CFT dual of the ``Little Bang"}

\author{Edward Shuryak\\
Department of Physics and Astronomy\\
University at Stony Brook,NY 11794 USA \\
shuryak@tonic.physics.sunysb.edu}
\maketitle
\begin{abstract}
This (rather subjective) review sums up few years of work devoted to
explain various aspects of high energy heavy ion collisions using the AdS/CFT
correspondence. The central issue of is is formation of the trapped surface (black hole)
phenomenon, seen by a distant observer as the entropy production.
We end up discussing an issue of classical gravitational radiation
by an ultrarelativistic falling body and the so called breaking self-force related to it.
\end{abstract}

\section{ Introduction}

 Spectacular conformation of the
hydrodynamical predictions at RHIC, as well as very strong jet quenching
(including heavy quarks) has lead to the ``strongly coupled Quark-luon Plasma" (sQGP) paradigm, see e.g. \cite{Shuryak:2003xe}. These events
put understanding of the strongly coupled regime of gauge theories phase at the front
of several communities involved. For a time, discussion of the main nonperturbative phenomena  -- confinement, chiral symmetry breaking, topology -- were put aside and
the community focused on  the deconfined/chirally high-T phase,
trying to understand the difference between the weakly and strongly coupled plasmas.

While such examples as strongly coupled electromagnetic plasmas and cold
fermionic gases at large scattering lengths has provided a lesson or two, 
a beautiful gift from the string community--  the AdS/CFT correspondence --
quickly became the major tool of the Nuclear Theorists, as well as a prime 
examples of the ``string theory applications".

Another major event in the field -- the first LHC heavy ion run, at the end of 2010 -- have produced nice confirmations of old hydrodynamical
predictions  and  even more spectacular display of jet quenching,
with up to a 100 GeV dissipated into the plasma. An example of
new theory development is ``the second act of hydro": the calculation
and observation of higher harmonics of the flow induced by initial fluctuations.
As an example of how useful a collaboration between two theory communities may be,
let me mention that behind our study of the higher flow harmonics \cite{Staig:2011wj}
(not to be discussed below) would not be possible without the zeroth order
conformal solution found by Gubser  \cite{Gubser:2010ze}.

  There is no need to go into detail into description of classic results obtained 
  in the AdS/CFT framework \cite{Maldacena:1997re}, such as
 thermodynamics of strongly coupled
\cal{N}=4 plasma \cite{Gubser:1998nz} and transport properties  derived
from the linearized hydrodynamics, resulting in the famous $\eta/s=1/4\pi$ expression for the viscosity-to-entropy ratio
\cite{Policastro:2002se} . Let me just mention another very significant achievement: a general
 derivation of the full nonlinear hydrodynamics with gradient expansion
 from the gradient expansion of the Einstein equation\footnote{In the 19th century 
 hydrodynamics was a very advanced theoretical field, teaching how to work with
 flows and sound waves. Stokes was clearly important for Maxwell, for example.  
 Yet in 1970's, when I was starting my studies of QGP
 and of hydrodynamics of high energy collisions, most theorists were telling me those have no chance, being ridiculously simplistic and  incompatible with quantum field theories.
 Apparently, not any more.
 }  
  \cite{Bhatt:2008jc,Bhatt:2008xc,Natsuume:2007ty}, based on the  ``membrane paradigm" for  the black hole horizon developed in 1980's.

  

This is a very subjective review, based mostly on works with my (then) graduate student Shu Lin, as we were working our way into the AdS/CFT territory. As the reader will see,
our main interests were related with the most difficult issues of QGP equilibration
and jet quenching, related wit strongly-out-of-equilibrium situations. Our main idea was  that equilibration in the AdS/CFT context has very simple intuitive meaning -- objects
 simply fall into the black hole because of gravity forces -- and the rest is just technical
 difficulties, which can be dealt with.  
    

\section{The worm-up: thinking about  the Maldacena dipoles}

Static pair of infinitely heavy fundamental quarks was the first ADS/CFT applications,
and Maldacena's dipole calculation is so well known to anyone, that a reader may wander why
do I want to write about it here. And yet, there are important lessons which are perhaps not
well appreciated, perhaps warranting to come back to it at some later time.
 
It is hardly necessary to reminding the setting, as the reader was probably teaching
it many times: a Nambu-Goto string 
 pends  due to gravity force into the 5-th dimension,
like in the famous catenary (chain) problem. The celebrated Maldacena's result, the {\em  Coulomb law at strong coupling} is
\be \label{eqn_new_Coulomb}
V(L)= -{4\pi^2  \over \Gamma(1/4)^4 }{\sqrt{\lambda} \over  L}
\label{coulomb}
\ee
contains
 $\sqrt{\lambda}$  instead of $\lambda$ in
the weak coupling. (The numerical coefficient in the first bracket is 0.228,
to be compared to the result from a diagrammatic
re-summation below.)

What is the reason for this modification? For pedagogical reasons
let me start with  ``naive  guesses"
:\\ (i) Perhaps the
 strongly coupled vacuum acts like some
kind of a space-independent dielectric constant, $\epsilon \sim 1/\sqrt{\lambda}$
which is reducing the effect of the Coulomb  field, but similarly at all points.\\
(ii) Perhaps such dielectric constant is nonlinear effects, and thus is
not the same at different points: but the fields created by static dipole
are still just the electric field $\vec E$.

 The AdS/CFT allows one to get many details about
 the problem and test such
ideas: but we cannot get fields directly.
Its rules allows one to 
find  $holograms$  on the boundary using bulk wave eqns
for massless  fields - which are scalars, vectors or gravitons. For example I
will give our results for boundary stress tensor obtained from
the (linearized) Einstein equation for the gravity perturbations
calculated in \cite{Lin:2007pv}, producing the {\em stress tensor}
of matter  $<T_{\mu\nu}(y)>$ at any  point $y$ on the boundary,
induced by the Maldacena dipole. 
  The solution is too technical to be presented here and even the
  resulting stress tensor expressions are too long. l
The leading terms  far from the dipole
$y>>L$ is
\ba\label{eq:ff}
T_{00}=\sqrt{\lambda} L^3
\left(\frac{C_1y_1^2+C_2 y^2}{|y|^9}\right)f(\theta)
\ea
where $C_1,C_2$ are numerical constants whose values can be looked up in the paper
and $f(\theta)$ is the angular distribution 
which is different that in  the weak coupling dipoles, for which
$T_{00}\sim \lambda L^2/y^6 $.
These and other results quickly put to rest simplistic ideas mentioned above:
one clearly create more fields than just the electric one, with two static charges.


  As we get glimpse of some first results from AdS/CFT  we see that they are quite different from weak coupling and we would like
    to understand them.
    In fact we would like to do so both from the bulk (gravity) side
as well as  from the
   {\em gauge theory} side.
   It turns out the first is relatively easy. For example, both in the total energy
   and energy density we get $\sqrt{\lambda}$ because this factor
   is in front of the Nambu-Goto Lagrangian (in proper units). The reason
   the field decays as $y^7$ is extremely natural:
    in the $AdS_5$ space the
   function which inverts the Laplacean (analogously to Coulomb
   $1/r$ in flat 3d) has that very power of distance
   \be P_s={15\over{4\pi}}\frac{z^2}{(z^2+r^2 )^{7\over2}}\ee
   with $z$ being the 5-th coordinate of the source and $r$ the 3-distance
   between it and the observation point\footnote{Please recall
   that in gravity there are no cancellations
   between different contributions: any energy source perturbs gravity with the same sign.}. 
   
    In order to understand the same results from the gauge side we will
    need a bit of pedagogical
   introduction: the resolution will be given by
   the idea  of {\em short color correlation time}
  by Zahed and myself \cite{Shuryak:2003ja}.
 In QCD, with its running coupling, 
 higher-order effects modify the zeroth order Coulomb field/potential. 
 Since we consider static problem, one can
rotate time into Euclidean time $\tau=it$, which 
not only makes possible lattice simulations but also 
simplifies perturbative diagrams.  
Let us do Feynman diagrams directly in the coordinate
representation. The lowest order energy, given by the diagram in which one
massless quantum (scalar or gauge component $A_0$) is emitted by one charge at time $\tau_1$ and absorbed by
another at time $\tau_2$  
is just 

\be V(L)(Time) \sim  -g^2\int_0^{Time} {d\tau_1 d\tau_2 \over L^2+
  (\tau_1-\tau_2)^2}\sim -{g^2 (Time)\over L} \ee
 where 'Time' is total integration time and
 the denominator -- the square of the 4-dim distance between
gluon emission and absorption  represents the Feynman propagator in Euclidean
space-time. The propagation time for a virtual quantum $(\tau_1-\tau_2)\sim L$,
thus the Coulomb $1/L$ in the potential.

 Higher order diagrams include self-coupling of gluons/scalars
and multiple interactions with the charges. A
famous simplification  proposed by 't Hooft is the large number of colors limit
in which only $planar$ diagrams should be considered. People suggested that
as $g$ grows those diagrams are becoming ``fishnets'' with smaller
and smaller holes, converging to a ``membrane'' or string worldline:
but although this idea
was fueling decades of studies trying to cast gauge theory into
stringy form it have not strictly speaking succeeded.
It may still be true: just nobody was smart enough to sum up 
all planar diagrams\footnote{Well, AdS/CFT is kind of a solution,
actually, but it is doing it indirectly.}.

  If one does not want to give up on re-summation idea, one may
consider a subset of those  -- the $ladders$ --
which can be  summed up.
Semenoff and Zarembo \cite{Semenoff:2002kk} have done that: let us look at what 
they have  found.
The first point is that in order that each rung of the ladder contributes a factor $N_c$,
 emission time ordering should be strictly
enforced, on each charge; let us call
these time moments $s_1>s_2>s_3...$ and $t_1>t_2>t_3...$.
Ladder diagrams must connect $s_1$ to $t_1$, etc, otherwise it is nonplanar
and subleading diagram.
 Thus the main difference from
the Abelian theory comes from the dynamics of the color charge itself!
The (re-summed) Bethe-Salpeter kernel $\Gamma(s,t)$, describing
the evolution from time zero to times $s,t$ at two lines,
satisfies the following integral equation
\be \label{eqn_BS}
\Gamma({\cal S},{\cal T})=1+{\lambda \over 4\pi^2}\int_0^{\cal S} ds
\int_0^{\cal T} dt{1\over
    (s-t)^2+L^2} \Gamma(s,t) \ee
If this eqn is solved, one gets  re-summation of all the
ladder diagram. The kernel obviously
satisfies the boundary condition $\Gamma ({\cal S},0) =\Gamma(0,{\cal T})=1$.
If the equation is solved, the ladder-generated potential is
\be
V_{\rm lad}(L) 
=-\lim_{T\to{+\infty}}{\frac 1{\cal T} \Gamma\, ({\cal T},{\cal T})}\,\,,
\label{0a}
\ee
In weak coupling $\Gamma\approx 1$  and the integral on the rhs is
easily taken, resulting in the usual Coulomb law. 
For solving it at any coupling, it is convenient
to switch to the differential equation

\be
\frac{\partial^2\Gamma}{\partial {\cal S}\,\partial {\cal T}} =
\,\frac{\lambda/4\pi^2}{({\cal S-T})^2+L^2}
\Gamma ({\cal S,T})\,\,\,.
\label{1a}
\ee
and change variables to
$x=({\cal S-T})/L$ and $y=({\cal S+T})/L$ through
\be
\Gamma (x,y) =\sum_{m}\,{\bf C}_m \gamma_m (x)\,e^{\omega_m y/2}
\label{2a}
\ee
with the corresponding boundary condition $\Gamma (x,|x|)=1$. The
dependence of the kernel $\Gamma$ on the relative times $x$ follows
from the differential equation

\be
\left(-\frac{d^2}{dx^2} -
\frac{\lambda/4\pi^2}{x^2+1} \right)
\,\gamma_m (x) = -\frac {\omega_m^2}{4}\,\gamma^m (x)
\label{3a}
\ee
For large $\lambda$ the dominant part of the potential in (\ref{3a})
is from {\it small} relative times $x$ resulting into a harmonic
equation~\cite{Semenoff:2002kk}

At large 
times ${\cal T}$,  the kernel is dominated by the lowest harmonic mode. For large times ${\cal S\approx T}$ that is small $x$ and large 
$y$ 

\be
\Gamma (x,y)\approx {\bf C}_0\,e^{-\sqrt{\lambda}\,x^2/4\pi}\,
e^{\sqrt{\lambda}\,y/2\pi}\,\,.
\label{5a}
\ee
From (\ref{0a}) it follows that
in the strong coupling limit the ladder generated potential
is \be V_{\rm lad}(L)= -\frac{\sqrt{\lambda}/\pi}L \ee which 
has {\em the same parametric form}  as the one derived from the
AdS/CFT correspondence (\ref{eqn_new_Coulomb}) except for the
overall coefficient. Note that the difference
is not so large,  since $1/\pi=0.318$ is larger than the exact value  
0.228 by about 1/3. 
Why did it happened that the potential is reduced relative to 
the Coulomb law by $1/\sqrt{\lambda}$? It is because the relative time
between gluon emissions is no longer $\sim L$, as in the Abelian case,
but reduced to parametrically small time of relative color coherence  $\tau_c\sim 1/L\lambda^{1/2}$. 

\begin{figure}[t!]
\includegraphics[width=7cm]{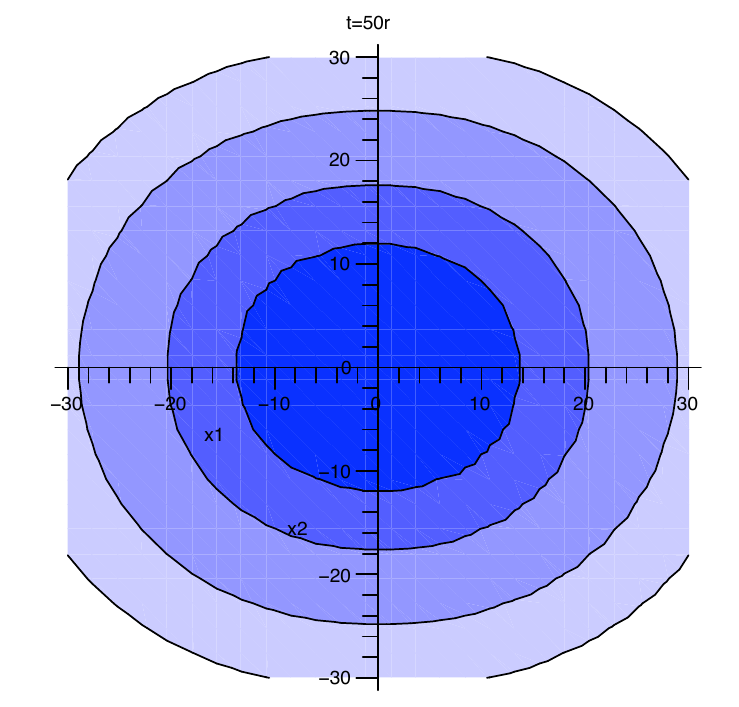}
\includegraphics[width=7cm]{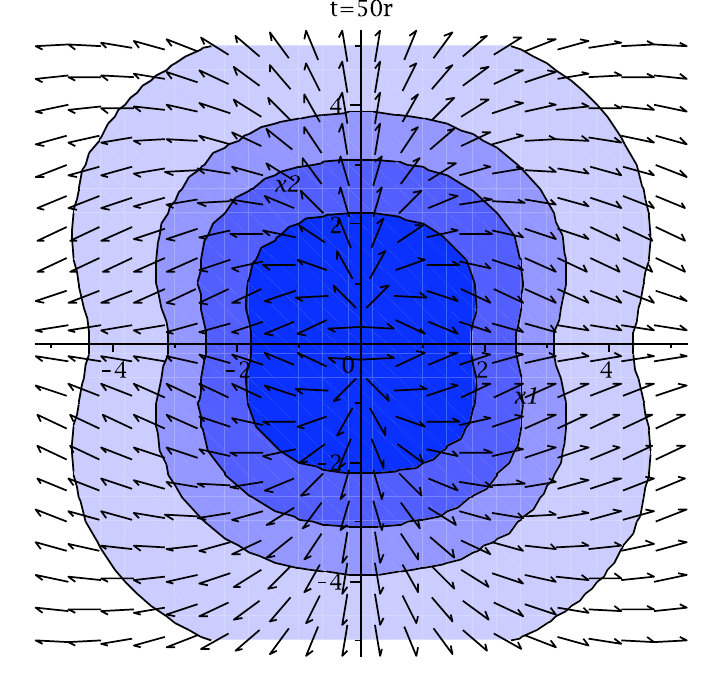}
 \caption{\label{fig_holo2} The hologram of a falling string.
The contours show the magnitude of the energy density (a) and the Poynting vector
$T^{0i}$ in the transverse plane (b) at some time.
The direction of the momentum flow is indicated by arrows.}
\end{figure}

\section{``Holographic $e^+e^-$ annihilation"  and the absence of the jets}

In 1976, three years after discovery of QCD, $e^+e^-$ annihilation experiments at SLAC
has seen the first indications for two jets, created by quark-antiquark pair moving away from each other. That lead to discussion of ``string breaking" phenomena, phenomenological models
of  onset of confinement in hadronic collision, known as the Lund model
(for 
B.Andersen and collaborators), whose descendants are used by experimentalists for this day.

 
  AdS/CFT version of the holographic $e^+e^-$ annihilation has been developed in two
works by S.Lin and myself \cite{Lin:2006rf,Lin:2007pv}. Before we describe some
of its results, let me mention another possible usage of those results.
Some of the dreams high energy theorists have about possible discoveries
at LHC or beyond is existence of ``hidden" interactions which we don't see
because the so called ``mediators" -- particles which are charged both
under Standard Model and hidden theory -- are heavy. Perhaps the hidden sector
is strongly coupled: if so one may wander how production of a pair
of new charges would look like. Further discussion of this issue (and  more
references) can be found in a paper by
Hofman and Maldacena\cite{Hofman:2008ar}, who made further
steps toward understanding the  ``strongly coupled collider physics''.

  The setting of
 Refs \cite{Lin:2006rf,Lin:2007pv} is simple:
two heavy quarks $departing$ from each other with velocities $\pm v$, on the boundary of the  AdS$_5$ space.   
The AdS/CFT require charges connected by a classical string which (unlike
the QCD one) : is not breaking but stretches indefinitely
 \footnote{This is true for classical string
minimizing the action. However
 account for fluctuations of the string allows it to touch
the flavor brane and break: in fact Peeters, Sonnenschein and Zamaklar
\protect\cite{Peeters:2005fq} have calculated
the holographic decays of large-spin mesons this way.}. 
 Before solving the corresponding equation in full, we will
 first discuss ``scaling'' solutions
in
the separable form 
\be z(\tau,\eta)={\tau\over f(\eta)}\ee
where $\tau,\eta$ are the proper time and space-time rapidity.
The corresponding solution was obtained, but
we found that it can
 only exists for  sufficiently small\footnote{In particular, for
velocities going to zero one finds a strongly coupled version of Ampere's
law, the interaction of two nonrelativistic currents.} velocities.
Moreover, the analysis of the
 classical stability of such scaling solution
revealed that it gets unstable  for $Y>Y_m\approx 1/4$, where
quark rapidity is related to their velocity in the usual way $v=tanh(Y)$.
For larger velocity of the quarks the scaling solution has to be
substituted by a non-scaling one, depending on both variables in a
nontrivial way, which was analyzed numerically. 

The physical picture an observer at the boundary would see
is again given by a 
 (gravitational) hologram of the falling string calculated in the second paper
\cite{Lin:2007pv}. One feature of the result should not be
surprising for the readers who followed the description of the
hologram of the $static$ Maldacena string above: indeed, no trace
of a string is in fact visible. The results, obtained after heavy numerical calculation --
 shown in
 fig.\ref{fig_holo2} -- produced a near-spherical explosion.
 This means that {\em there are no jets at strong coupling}! 
 
 Furthermore
  we found this explosion to be {\em non-thermal and
thus non-hydrodynamical}, in the sense that the stress tensor
found (although of course conserved and traceless)
cannot be parameterized by the energy density and isotropic
pressure. Indeed, thermal physics would require a black hole which is not at hand.

\begin{figure}[t]
\includegraphics[width=0.5\textwidth]{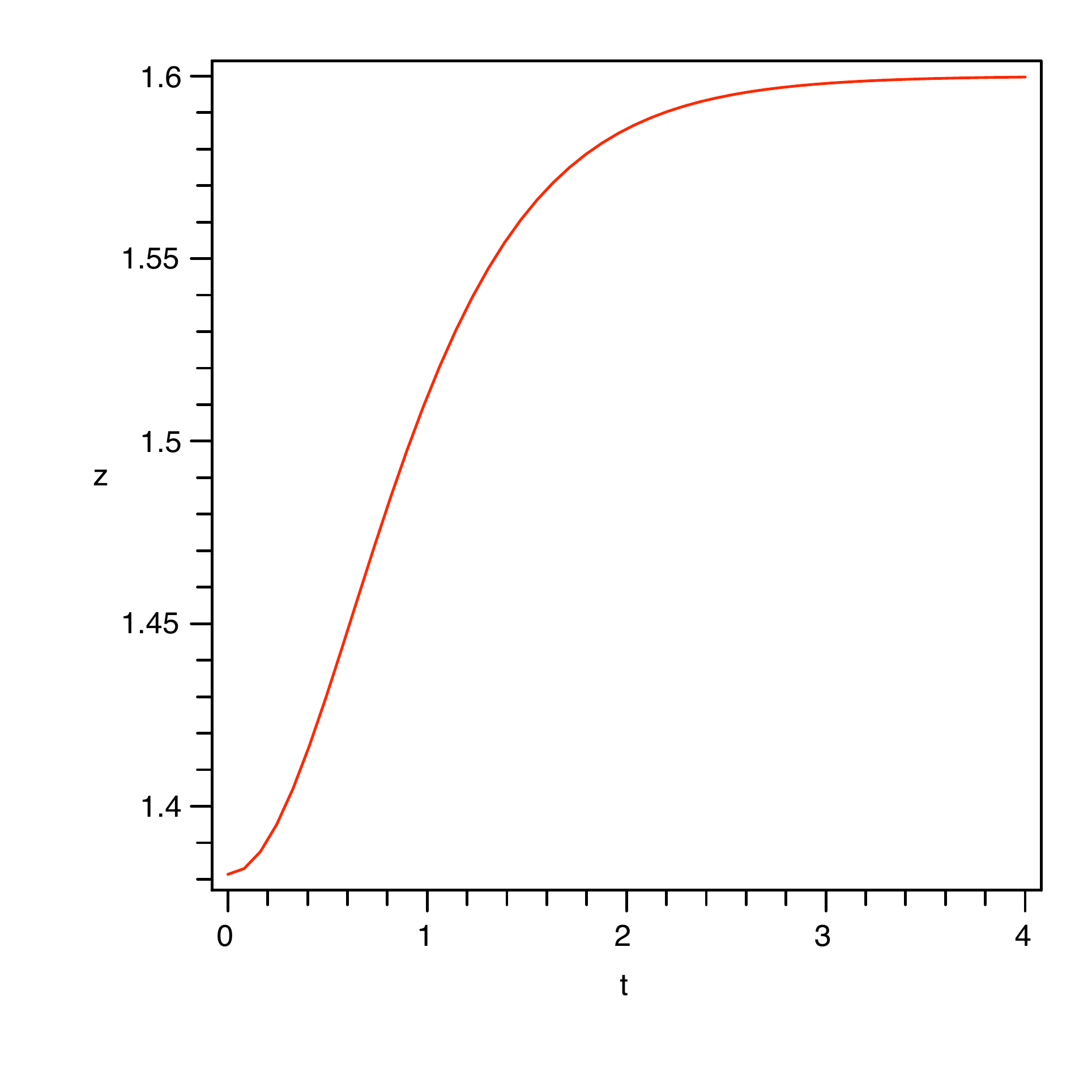}
\vspace{-.5cm} 
\includegraphics[width=0.47\textwidth]{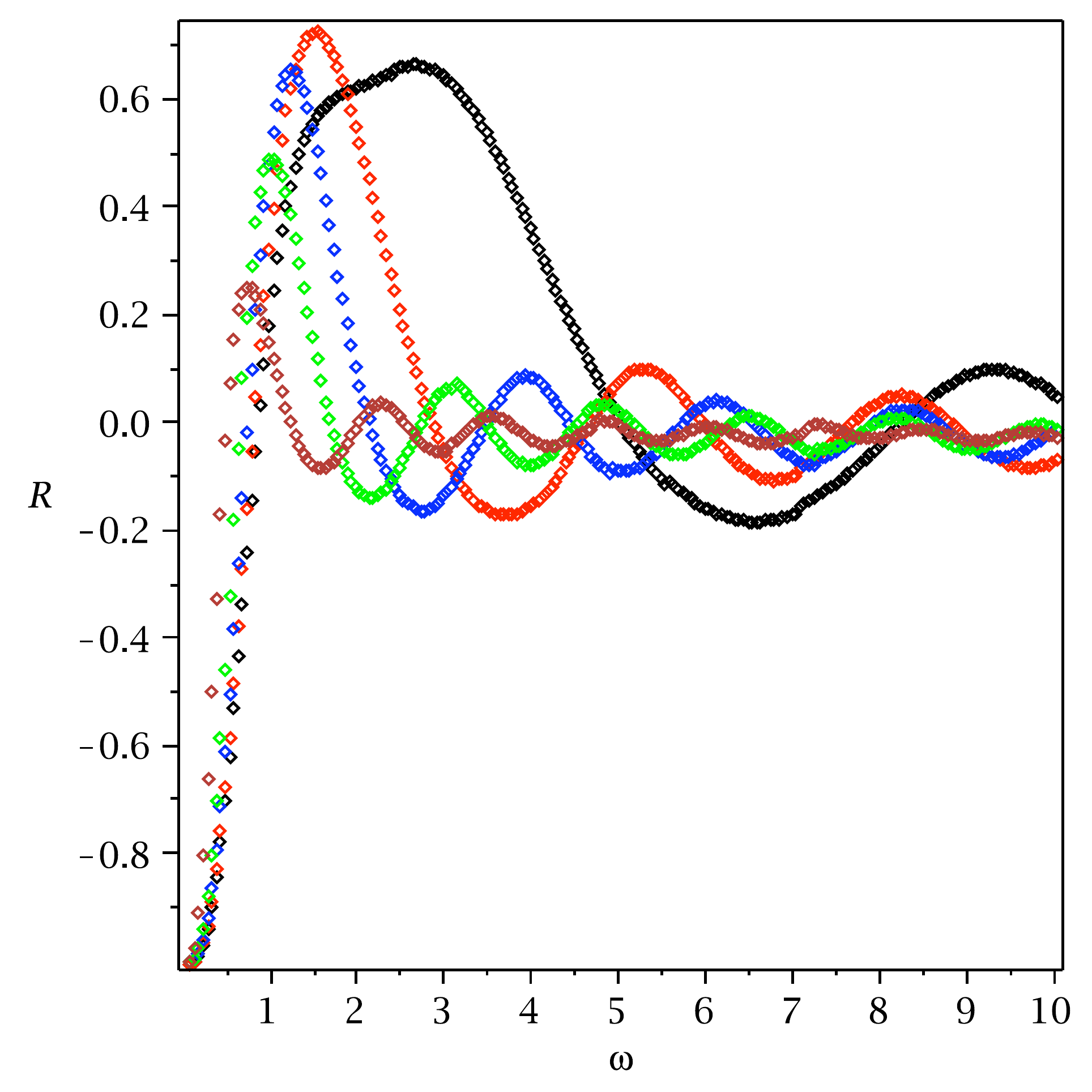}
\caption{\label{stages} (a) The shell trajectory as a function of time.
It starts at rest at $z=z_0$
with a constant acceleration, followed by a constant falling and 
eventually approaches the horizon in a exponential fashion. 
The parameter we choose are $\kappa_5^2 p=1$ and $z_h=1.6$.
(b)The relative deviation $R$ of the spectral density 
at zero momentum $q=0$ at different stages of 
thermalization.  The oscillations are explained by ``echo" effect. }
\end{figure}
\section{ ``Collapsing membrane" and the dynamics \\ of the equilibration scale }

  The string is  only one type of ``bulk" objects which may fall into the AdS center. 
In relatively early paper
 Sin, Zahed and myself~\cite{Shuryak:2005ia}
first argued that exploding/cooling
fireball, created in high energy collisions,  is  a
  black hole 
  formed by the collision debris and then 
falling toward the AdS center. While the specific solution 
 discussed in that paper was spherically symmetric and thus  more 
 appropriate for cosmology than for heavy ion
applications, it was a direct predecessor of series of papers by Janik
and collaborators on ``rapidity independent" falling horizons.

    In this section we would start discussing the issue of equilibration,
    keeping to a picture that UV modes are fast and equilibrate quickly,
    while in IR the life is more slow and equilibration is delayed.
    Thus the idea of certain ``equilibration front" appears, moving in scale from UV to IR.
  In AdS/CFT language this front can be thought of like some material objects,
  falling along the ``scale" coordinate $z$.
     
The paper in which it has been worked out , by
Lin and myself \cite{Lin:2008rw}, considered the simplest geometry
in which  a shell (elastic membrane) is flat in $x_1,x_2,x_3$ and is collapsing 
{\em under its own weight} in AdS$_5$. 
The metric is simple: it is thermal  AdS above the shell and empty AdS below.
The only questions are what is the equation of motion of the shell and what
can different observers see on the boundary (in the gauge theory).

An elegant element of the paper
is the systematic use of the so called Israel junction condition.
The falling velocity (in time of the  distant observer $t$) is given by:
\be\label{trajectory}
\frac{dz}{dt}=\frac{\dot{z}}{\dot{t}}
=\frac{f\sqrt{(\frac{\kappa_5^2 p}{6})^2+(\frac{3}{2\kappa_5^2 p})^2(1-f)^2-\frac{1+f}{2}}}{\frac{\kappa_5^2 p}{6}+\frac{3}{2\kappa_5^2 p}(1-f)}
\ee
where $\kappa_5^2 p$ is 5-dim gravity constant and shell elastic constant and 
$f=1-z^4/z_h^4$ as usual, in thermal AdS. The value of the horizon position $z_h$,
to which the shell asymptotically goes, see Fig.\ref{stages} depends on the total mass of the shell (as detailed in the paper).

The main question is by which experiments an observer on the
boundary can distinguish the true thermal state (thermal AdS metric)
 from ``quasiequilibrium'' one in our solution. 
A ``one-point observer" would simply see the
equilibrium pressure and energy density, as the metric above the shell is thermal AdS.
Yet more sophisticated 
``two-point observers"  can measure correlation functions, and they will see
deviations from  the equilibrium because a signal can penetrate $below$ the shell. Solving for various two-point functions
in the background with falling shell/membrane we
found such deviations: they
are oscillating in frequency around thermal ones. The corresponding
oscillation is explained by certain ``echo" times for  a signal reflected from the shell:
see more on this interesting phenomenon in  \cite{Lin:2008rw}.

\section{Entropy production and the trapped surfaces   }

  Multiple attempts over the years to understand the
 initial equilibration of QGP and provide  quantitative theory of the entropy production
 has not yet succeeded. The so called 
{\em ``glasma''} approach, treating
random  gluonic fields 
via classical Yang-Mills equations, has been pursued by many groups.
Related phenomenology tells us that
the typical momenta of those gluons, known as the
``saturation scale''  $Q_s$, is at RHIC/LHC in the range of 1.5/2 GeV. This is uncomfortably close
to their  momenta in the sQGP at hydrodynamical stage to follow, the ``most perfect liquid" as we now know, and 
 one
may wander if the perturbative (weak coupling) approach currently used
is indeed justified.

General relativity 
provides
  a  remarkably simple mechanism of the entropy production:
 as the information happened to be surrounded by the
 ``trapped surface" cannot be retrieved, it is  the entropy produced!  Even more
 convenient fact is that the trapped surfaces do not respect the usual causality arguments:
 they can form in the perfectly empty space, in $anticipation$ of the collision to follow later, as they are always justified {\em a posteriori}.  This leads to a paradoxical conclusion of some entropy
 being there already at t=0, as
 one indeed finds the trapped surface being there already. 
 

\begin{figure}[t]
 \includegraphics[width=10cm]{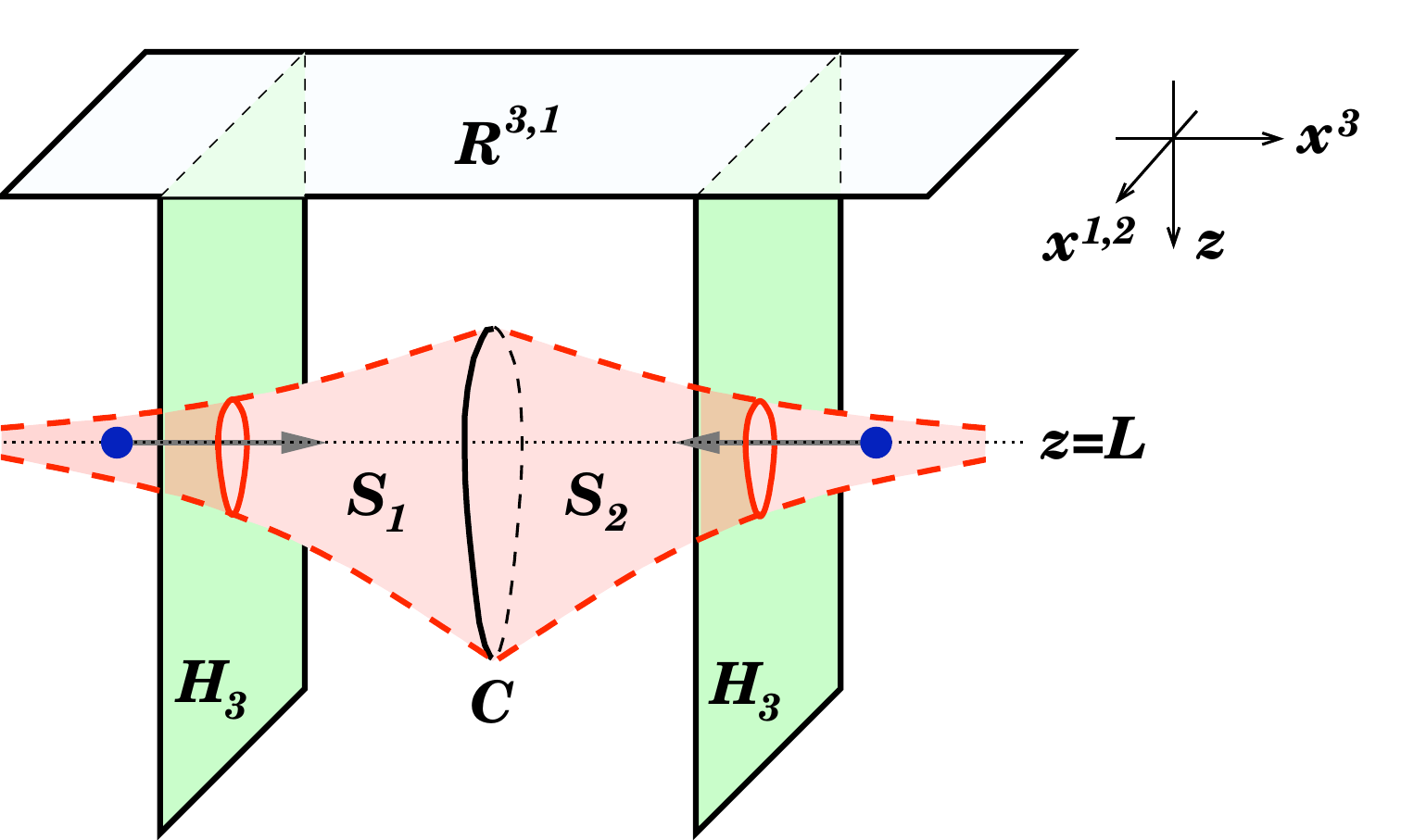}
 \includegraphics[width=6cm]{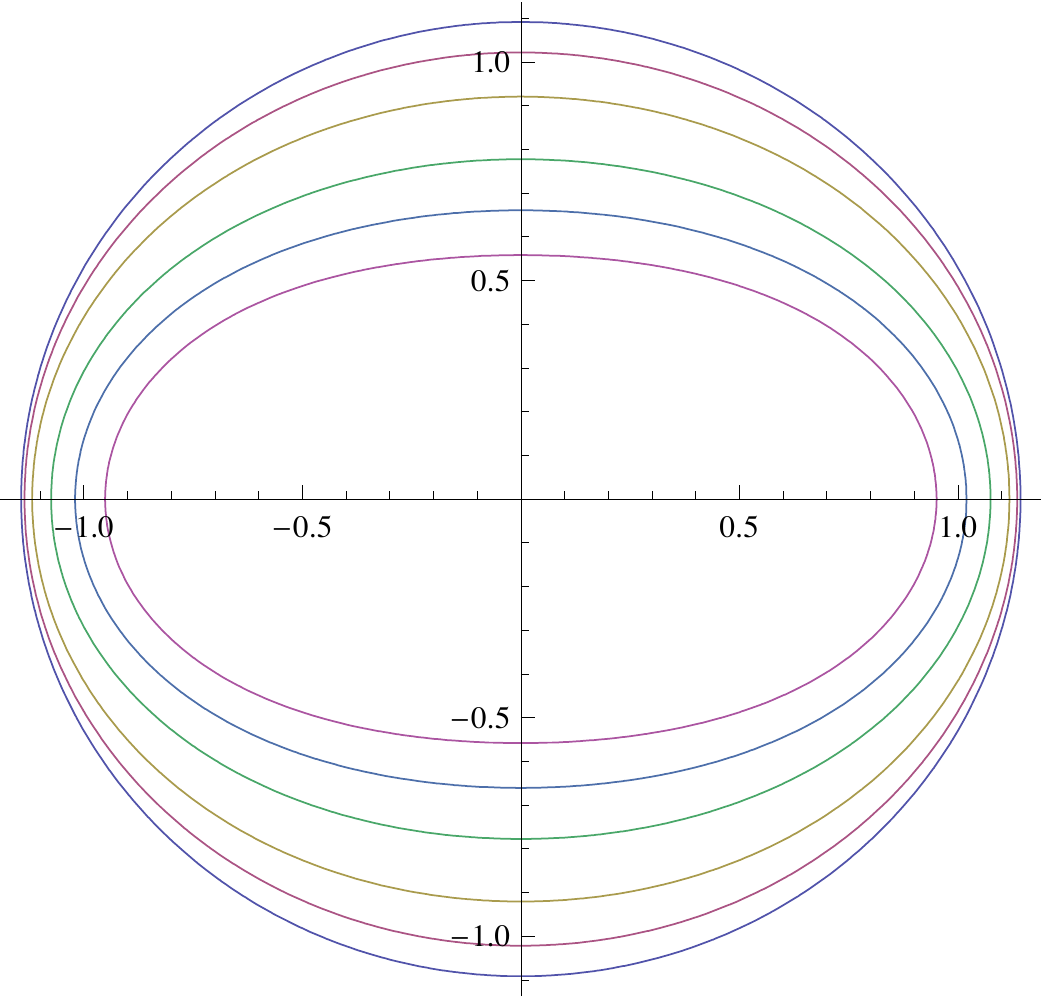}
\caption{(a) From \protect\cite{Gubser:2008pc}:A projection of the marginally trapped surface that we use onto a fixed time slice of the AdS geometry.  The size of the trapped surface is controlled by the energy of the massless particles that generate the shock waves.  These particles are shown as dark blue dots.
(b) From \protect\cite{Lin:2009pn}.The shapes of 
the trapped surface at at collision time in the transverse plane, for impact parameters  $0.4L,\,0.6L,\,0.8L,\,
1.0L,\,1.1L,\,1.14L$ from the outer to the inner, the innermost 
being the critical trapped surface.
} 
\label{fig_trapped}
\end{figure}

 The first step  had been done  by Gubser, Pufu and Yarom           
\cite{Gubser:2008pc}, who proposed to look at heavy ion collision as a
process of head-on collision of two point-like black holes, separated from the
boundary by some depth $L$ 
see Fig.\ref{fig_trapped}(a).  By using global AdS coordinates,
these authors argued that (apart of obvious axial O(2)
symmetry) this case  has higher -- namely O(3)-- symmetry with the resulting
black hole at the collision moment at its center,
thus  in certain coordinate\be q={\vec x_\perp^2+(z-L)^2\over 4 z L } \ee
 the 3-d trapped surface C at the collision
moment should be just a 3-sphere, with the radius $q_c$. 
(Here $x_\perp$ are two coordinates transverse to the collision axes.)
The picture of the surface, taken from their paper, is shown in Fig.\ref{wigglyhor}.
One can find $q_c$ 
and determine its relation to CM collision energy and
Bekenstein entropy. For large $q_c$ these expressions are just 
\be E\approx {4 L^2 q_c^3 \over G_5}, \hspace{.2cm} S\approx {4 \pi L^3 q_c^2 \over G_5}, \ee
from which, eliminating $q_c$, the main result of the paper follows, namely that
the entropy grows with the collision energy as
\be S\sim E^{2/3} \label{eqn_Gubser_S}\ee
Note that this power is in general $(d-3)/(d-2)$  directly relates to the $d$=5-dimensional gravity, and is different
from the 1950's prediction of Fermi/Landau who predicted the power 
1/2. This prediction is also not far from reality\footnote{Simplistic comparison with total multilicity ignores the 
nonzero baryon number, important especially at lower collision energies,
which is not produced and should be removed.}


The generalization to non-central collisions has been done by Lin and  myself
 \cite{Lin:2009pn}. There is no
point in going into technical details here, so let me describe the equations in words. They resemble
electrostatic problem with two charges inside the metallic cavity, except that there is an extra
boundary condition on the fields (from smoothness of the surface) and, last but not least,
the surface is not given but has to be determined.  The way to solve the problem was via
rewriting it as integral equation. 

\begin{figure}[t!]
	\includegraphics[width=8.cm]{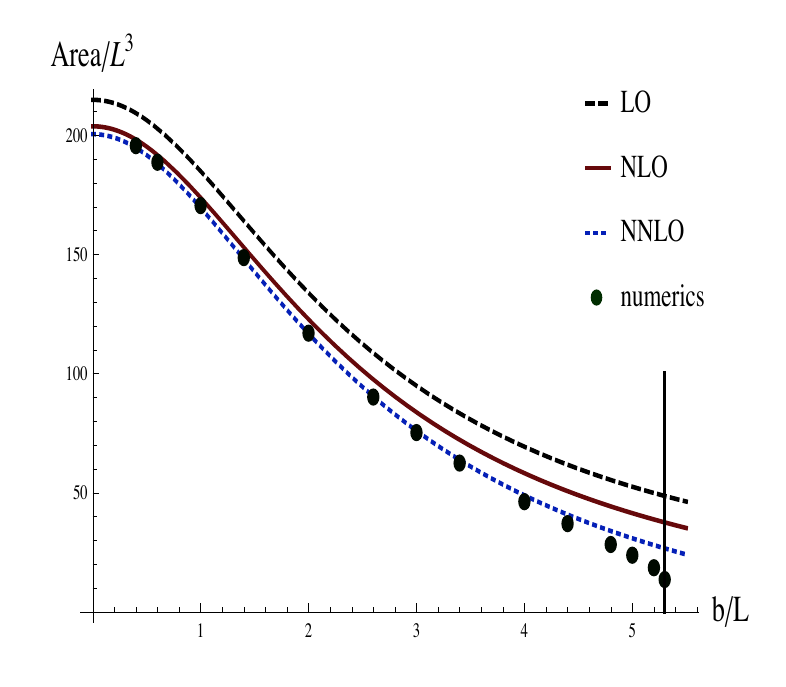} 
	\includegraphics[width=9.cm]{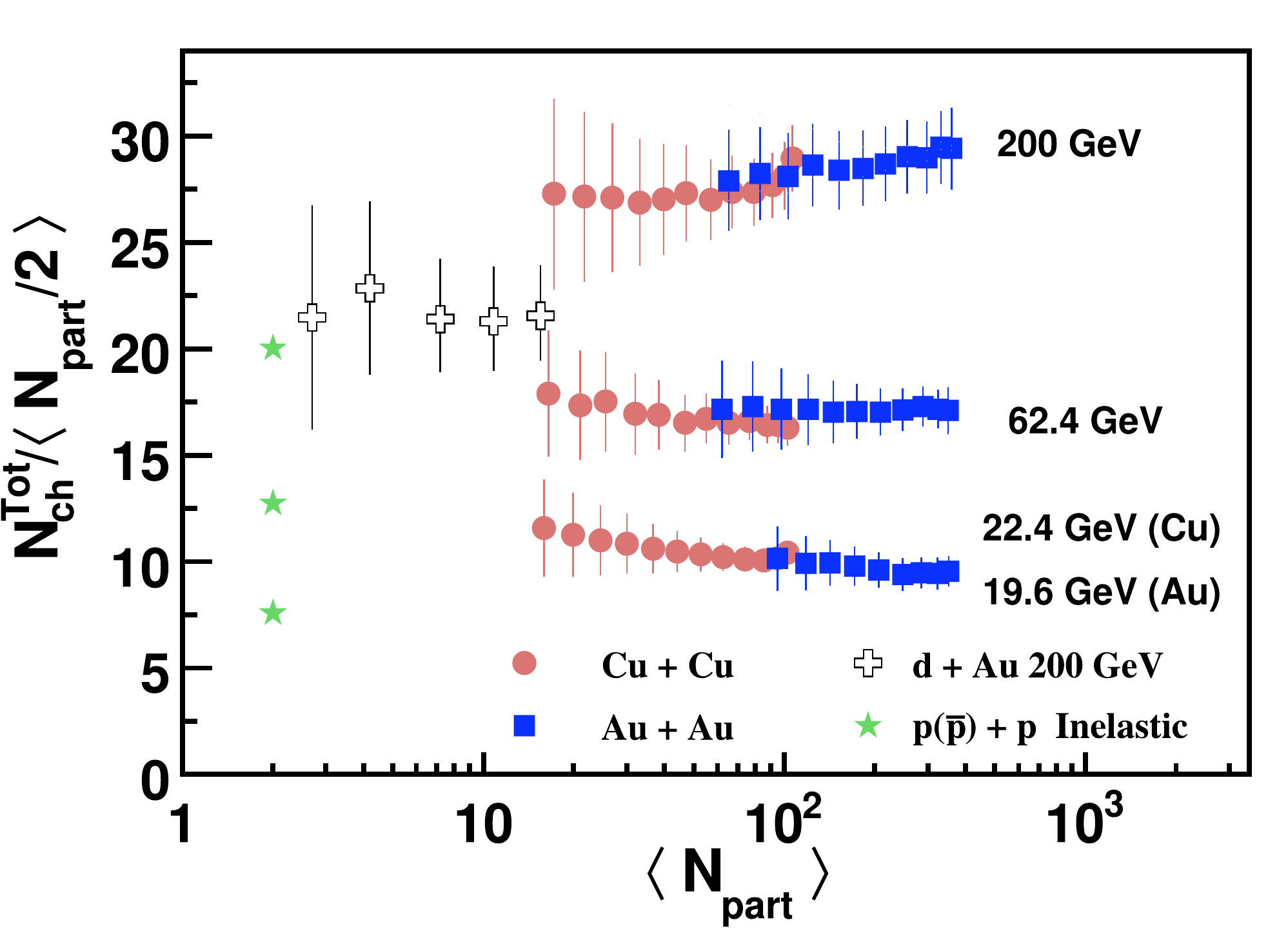}
\caption{ (a) The area of the trapped surface versus impact parameter, with the comparison of the numerical studies by Lin and myself \protect\cite{Lin:2009pn} shown by points and
analytic curves from \protect\cite{Gubser:2009sx}.
(b) Experimental data from PHOBOS collaboration, on the charge particle multiplicity at RHIC per participant nucleon. Three sets of data, one above the other, are for three collision energies, 200, 62.4 and 19.6/22.4 GeV. All three show  values independent on centrality of the collision (left peripheral, right central) for AuAu collisions (closed points), which is different and higher if compared to pp collisions
(green stars) and dAu collisions (open squares) in which no QGP is formed.  
}
\label{fig:noncentral}
\end{figure}

From gravity point of view the qualitative trand was clear: two colliding objects may merge
into a common black hole only provided that the impact parameter is less than some critical value $b_c(E)$, depending on the collision energy. Indeed, with $b$ rising, the trapped energy decreases and
angular momentum increases, so at some point no black hole can be formed. 
Interestingly, it happens as a jump, just a bit below this impact parameter reasonable trapped surface and black hole exist and nothing obvious indicates that at large $b$ none is formed.
The shape of the trapped surface is shown in Fig.\ref{fig_trapped}(b), and the
dependence of the trapped surface area on impact parameter is shown in Fig.\ref{fig:noncentral},
from next paper, comparing our points with a series of curves later obtained  by Gubser, Pufu and Yarom \cite{Gubser:2009sx}.

From the point of view of the gauge theory this behavior is however a complete surprise: it predicts first order transition as a function
of impact parameter, with creation of thermal fireball with significant entropy at $b<b_c$, while no such things is there for $b>b_c$. (Needless to say, classical gravity and such jumps are consequences
of the large $N_c$ approximation, presumably smoothened out at finite $N_c$.)
As discussed in our paper  \cite{Lin:2009pn}, the  experimental multiplicity data from PHOBOS
experiment at RHIC shown in Fig. suggest a relatively sharp transition from ``heavy-ion-like" to ``pp-like" entropy
as a function of impact parameter, but the experiments have not yet addressed the transition
from one to another well enough, to make this comparison quantitative.

In the same paper  \cite{Lin:2009pn} we have pointed out that the simplest
geometry to study the trapped surface would be wall-wall collision, in which
there is no dependence on transverse coordinates $x^2,x^3$ and thus a sphere becomes just two points in $z$, above and below the colliding bulk objects.
We elaborated on this in more recent work \cite{Lin:2010cb}, considering
collision of two walls made of material with different ``saturation scales"
(e.g. made of lead and cotton) and studied conditions for trapped surface formation.

(In connection with this, let us note that some papers consider collisions of walls
which have no source in the bulk and are $\sim \delta(x\pm t) z^4$ perturbation
of the metric. Although such walls may have the same boundary hologram as ours,
they lead to mathematical inconsistencies. They cannot be treated as 
perturbations at large $z$, and their trapped surfaces are not closed from below,
and thus their area is not even defined.) 

\section{The resummed hydrodynamics and the entropy at LHC}

Not all the entropy is produced promptly: some is produced 
by the dissipative effects  at the hydrodynamical stage.
In order to estimate this amount one first need  to specify what exactly is meant  by ``hydro"  and by its ``start".  To define a starting moment 
 is relatively easy:   any theory  may be considered  valid as long as it works with some preset accuracy (say,  one percent).
hydro itself comes at least in four forms: (i) ``ideal hydrodynamics"
without dissipation, (ii) Navier-Stokes  (NS) which includes the viscous term, (iii) the ``second order" hydro with the second gradients of some kind; (iv) the ``resummed" hydro suggested at least for
 AdS/CFT in \cite{Lublinsky:2007mm}).   
 
 \begin{figure}[t]
\centerline{ \includegraphics[width=10cm]{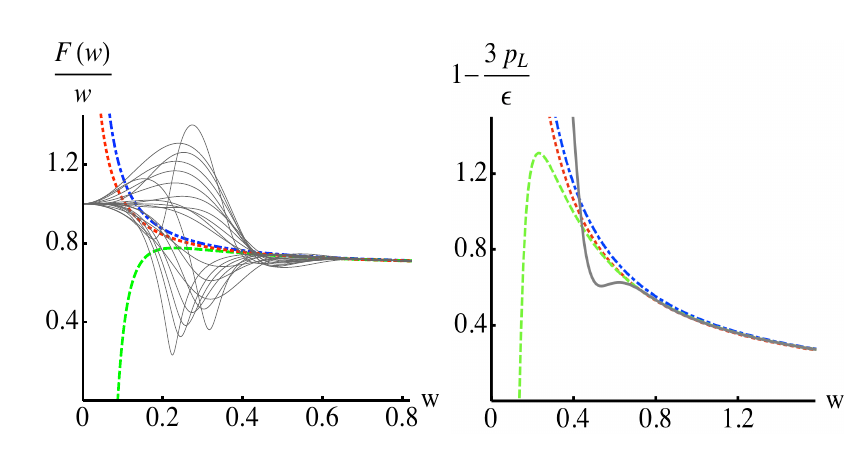}}
\caption{(left) The temperature evolution combination $dlog(w)/dlog\tau$ for different initial conditions (black thin curves) converging into a universal function of $w=T\tau$, compared to hydro. (right) The pressure anisotropy
for one of the evolutions compared to 1-st (NS), 2-nd and 3-ed order hydrodynamics.
} 
\label{fig_equilibration}
\end{figure}

 On a theory side, there has been a significant progress in solving
 Einstein equation for ``gravitational collisions", see e.g. \cite{Chesler:2010bi}. We will not go into this
 vast subject here, and only focus on the central issue of equilibration
 and onset of hydrodynamics. A very interesting study has been recently 
performed in Ref. \cite{janik2}. In the rapidity-independent approximation,
these authors had follow evolution for $\sim 20$ different 
and arbitrary initial conditions, and study how they equilibrate.
 Fig.\ref{fig_equilibration}(left), from this paper, show convergence of
 all of those evolutions to some universal function of the variable $w=\tau\,T$
\be  {d w\over d\ln \tau}\,=\,F(w)\,,\ee
whose existence is the essence of the ``resummed hydro" proposed in \cite{Lublinsky:2007mm}). As one can see, depending on accuracy, on may
assign the beginning of hydro to some ``initial" $w_i=0.4..-.6$. The plot on the right
demonstrate that at such time the anisotropy is still large and viscosity is important. 

The issue got into focus recently due to one of the first discoveries made at the first LHC  PbPb run. Indeed, it was found
 \cite{Aamodt:2010pb} that  the multiplicity in PbPb
collisions grow with energy more rapidly than  in pp:
\be
{dN^{PbPb}_{ch}\over dy}(y=0,s)\sim s^{0.15}  \hspace{1cm}  {dN^{pp}\_{ch}\over dy}(y=0,s)\sim s^{0.11} 
\ee
and the absolute value is 30-40\% larger than the CGC estmates calibrated by pp data. In other words, From the RHIC energy ($E=0.2\, TeV$) to the LHC, the double ratio is 
  \be \label{dr}
  {{dN\over d\eta}|_{PbPb,LHC}\,\,/\,\, {dN\over d\eta}|_{pp,LHC} \over  {dN\over d\eta}|_{AuAu,RHIC}\,\,/\,\, {dN\over d\eta}|_{pp,RHIC}} = 1.23\,. \ee
This  noticeable change with the energy  calls for a theoretical explanation.

Lublinsky and myself \cite{Lublinsky:2011cw} recently address both issues. In short,
we propose a simple form for the function $F(w)$ and calculate the entropy produced, from the time $w_i$ on. It turns out to be about 30\%.
Furthermore, we get the following expression for the contribution
to this double ratio  $\approx 1+{3[\bar\eta(LHC)-\bar\eta(RHIC)]\over  2w_i+3\bar\eta(RHIC)} $ and show, that the observed growth can be naturally explained by the viscosity growth, from RHIC to LHC, predicted by a number of phenomenological models.

\section{Jet quenching and the longitudinal ``self-force"}

There is no place here to review the well known works which established jet quenching
for heavy  and light  quark jets: it is clearly done in other reviews of this volume. Focusing on the latter case, it is sufficient to say
that those basically describe the quenching process by falling of some massive object in radial AdS
coordinate, toward the AdS center (or into IR). ``Quenching" is not seen in the bulk, as 4-momentum
of the falling object is conserved, but the holographic pictures at the boundary do show Mach cones
in very good agreement with  hydro predictions. 

 This article is also not a place to provide a review of jet quenching phenomenology. Let me just comment that
RHIC reconstructed jets and especially dijet LHC data from the first PbPb run have significantly changed the views of
its mechanism. Huge losses of the associate jets, reaching $O(100\, GeV)$, together with basically unchanged
narrow distribution in jet relative angle at $\pi$, point to robust longitudinal breaking force rather than transverse kicks
predicted perturbatively. Its magnitude apparently way exceed the tensions of the QCD string ,    
and probably even that of the AdS strings on which the abovementioned heavy quark quenching theory has been based.
  
  So, what may the physical nature of this longitudinal force be? 
  Recent  work by Yee, Zahed and myself \cite{Shuryak:2011ge}
 suggests to look at
  next-order bulk processes including graviton/dilaton radiation of the ``falling body".  Its non-AdS (weak coupling) predecessors
  includes my 1973 paper with Khriplovich \cite{Khriplovich:1973bq} in which electromagnetic and gravitational radiation for ultrarelativistic
  particle in gravity background has been calculated, as well as  the 2002 paper with Zahed \cite{Shuryak:2002ai} in which
  we discussed classical QCD syncroton-like radiation in transverse color field.  
  
  The calculation of gravitational radiation from an ultrarelativistic particle moving in a curved background is a notoriously difficult problem. 
 However, a local version of the expression for self-force has been
 derived  \cite{Quinn:1996am,Mino:1996nk} who,
 in gravittional setting,  obtained a remarkable contribution to  the ``self-force"
  \ba a^a = m u^b u^c \int_{-\infty}^{\tau^-}d\tau'  u^{a'} u^{b'} ( {1\over 2} \nabla^a G_{bca'b'} 
 - \nabla_b G_{c\,\,a'b'}^{\,\,a} -{1 \over 2} u^a u^d \nabla_d G_{bca'b'}
 ) \ea 
 in which the integral is done over proper time and the past  world line of the particle till
 regulated present time $\tau^-$, and $G$ is the retarded Green function for the Einstein equation
 with the particle as the source.
 Note that the bracket is just the Chrystoffel force for a gravity perturbations, induced by the past history of the particle itself. 

  Returning to electrodynamics in flat 3+1 dimensional space
  one finds that there is no such local expression: and furthermore,
  there cannot be, as the retarded propagator   
  is $\sim \delta(x^2)$ and localized on the light cone, never
  intersecting the particle path in the past. Yet we observed that
  it is not true in odd space-time dimensions, and calculated (in a separate
  companion paper) the radiation and self-force in 2+1 and 4+1 space-times: the results match the formula. 
  
  Encouraged by it, we calculated the self-force using the small-time expansion of the retarded propagator. 
  Many structures vanish because covariant acceleration (and higher derivatives)
  are zero on the geodesic. Also thermal AdS has simplified Ricci tensor,
  proportional to the metric itself, thus $R_{ij}\dot{x}^i\dot{x}^i\sim \gamma^0$,
  not the square of the jet gamma factor $\gamma^2$ as one would naively expect. But the quadratic Riemann term
  is such that all indices of the 4 velocities generate the maximal power of $\gamma$.
  The resulting breaking force
  \begin{eqnarray}
m{\ddot{x}}^a\approx-\frac{ G_5 m^2}{30\pi}\,\left(\int d\epsilon\right)
\,{{{{\bf R}^m}_{e}}^{n}}_{b}{\bf R}_{mcnd}\,\,
\dot{x}^e\dot{x}^b\dot{x}^c\dot{x}^d\,\,\dot{x}^a,\nonumber\\
\label{SELF}
\end{eqnarray}
  where the small parameter $\epsilon \sim 1/\gamma^2$ due to acceleration-induced 
  curving of the path.  The relativistic force  thus
is\footnote{Thus the usual non-relativistic force is $O(\gamma^2)$,
same power as in Schwartzschield metric  \cite{Khriplovich:1973bq}.} 
 $\sim \gamma^3$. The 5-dim gravitational constant is small,
 $G_5\sim 1/N_c^2$ yet power of the jet gamma factor 
 is large. Not only this self-force seem to be very large, it is also
  rapidly grow with the distance travel, stopping the jet earlier than
  the geodesic fall would predict. 
 The issue of self-force and its relation  to the gravitational radiation
in thermal AdS are rather new, and should of course be studied further,
before ay phenomenological applications can be discussed. 
 
\section{Summary}
While AdS/CFT correspondence does not hold for QCD and it is not good to predict particular numbers, it clearly is
hugely successful in
providing correct qualitative picture of strongly coupled dynamics. It is also constantly
 forcing the theorists to look at the familiar problems in a completely new light. Naturally, as we
learn more and more of it, the problem discussed are becoming more complex. Many efforts
are going into more and more ``realistic" solutions for ultrarelativistic gravitational collisions,
the dual to the ``Little Bang" we hope to find.

Let me end by
reiterating few applications/conclusions/questions discussed above.

{\em ``Maldacena dipole"} is next simplest field configuration, after that of one charge. And while
we know scalar and stress tensor picture of it, we still have no idea what fields are involved
and what shape they have. Early works on trying to resum certain diagrams to reproduce
the strong coupling answers were made but not followed. 

{\em ``No jets in $e^+e^-$ at strong coupling "}:  heavy quarks are always connected by the string, but its falling  into the 5-th dimension nontrivially depend on the velocity, and it is so rapid
that there is no trace of a string left in the resulting hologram on the boundary.   

{\em  ``Collapsing membrane"}  describes time-independent relaxation process,
in which relaxation proceed from UV to IR modes. Elegantly it 
 is described by a membrane falling into IR under its own weight.
The scale , being the $z$ position of the membrane,  changes as
prescribed by the  equation following from Israel junction condition. While ``single-point-observer" finds matter
apparently thermal, the ``two-point-observed" are able to look under the membrane and find the non-equilibrium modifications.

{\em ``Trapped surfaces"} can be instantly  formed in the collisions, providing
 estimates (from below) of the produced entropy. Remarkably, as a function of impact
parameter the solution disappear abruptly. This correlates well with relatively rapid switch from ``no hydro" to ``full hydro"
regimes in experiment, as the impact parameter varies.    

{\em ``Universal hydro" regime} can be studied by colliding objects with various initial
distributions in scales. We gave arguments that LHC entropy growth in PbPb collisions
may be due to anticipated changes in viscosity. 

{\em Gravitational radiation self-force} is new and
 rather difficult thing to understand/calculate.  It remains unknown how 
 our estimates correlate with  the actual intensity of bulk
gravitational radiation. Yet it seems clear that effects which are formally suppressed by power of $N_c$ can still be 
perhaps dominant in practice, due to very  large jet Lorentz factor.

But truly important would be to build a bridge between weak and strong coupling limits, which look at the moment disconnected.
This is why I included our discussion of Coulomb resummation,
Another fascinating possibility is establishment of the correspondence between fantastic progress in perturbative analysis of $\cal N$=4
scattering amplitude and their strong coupling limits, derivable from AdS/CFT. 

\section{Acknowledgements}
  My understanding of AdS/CFT applications, at whatever depth it is, came from multiple
discussion with many people, but mostly with my Stony Brook collaborators
Ismail Zahed,  Shu Lin and Derek Teaney.
 The work is partially
supported by the US-DOE grant 
DE-FG03-97ER4014.


\begin{thebibliography}{99}



\bibitem{Shuryak:2003xe}
  E.~Shuryak,
  Prog.\ Part.\ Nucl.\ Phys.\  {\bf 53}, 273 (2004)
  [arXiv:hep-ph/0312227].

\bibitem{Staig:2011wj}
  P.~Staig and E.~Shuryak,
  arXiv:1105.0676 [nucl-th].

\bibitem{Gubser:2010ze}
  S.~S.~Gubser,
  Phys.\ Rev.\  D {\bf 82}, 085027 (2010)
  [arXiv:1006.0006 [hep-th]].

\bibitem{Maldacena:1997re}
  J.~M.~Maldacena,
  Adv.\ Theor.\ Math.\ Phys.\  {\bf 2}, 231 (1998)
  [Int.\ J.\ Theor.\ Phys.\  {\bf 38}, 1113 (1999)]
  [arXiv:hep-th/9711200].

\bibitem{Gubser:1998nz}
  S.~S.~Gubser, I.~R.~Klebanov and A.~A.~Tseytlin,
  Nucl.\ Phys.\  B {\bf 534}, 202 (1998)
  [arXiv:hep-th/9805156].

\bibitem{Policastro:2002se}
  G.~Policastro, D.~T.~Son and A.~O.~Starinets,
  JHEP {\bf 0209}, 043 (2002)
  [arXiv:hep-th/0205052].



 \bibitem{Bhatt:2008jc}
  S.~Bhattacharyya, V.~E.~Hubeny, S.~Minwalla and M.~Rangamani,
  JHEP {\bf 0802}, 045 (2008)
  [arXiv:0712.2456 [hep-th]].
\bibitem{Bhatt:2008xc}
  S.~Bhattacharyya {\it et al.},
  arXiv:0803.2526 [hep-th].


\bibitem{Natsuume:2007ty}
  M.~Natsuume and T.~Okamura,
  Phys.\ Rev.\  D {\bf 77}, 066014 (2008)
  [Erratum-ibid.\  D {\bf 78}, 089902 (2008)]
  [arXiv:0712.2916 [hep-th]].

\bibitem{Lin:2007pv}
  S.~Lin and E.~Shuryak,
  Phys.\ Rev.\  D {\bf 76}, 085014 (2007)
  [arXiv:0707.3135 [hep-th]].

\bibitem{Shuryak:2003ja}
  E.~Shuryak and I.~Zahed,
  Phys.\ Rev.\  D {\bf 69}, 046005 (2004)
  [arXiv:hep-th/0308073].

\bibitem{Semenoff:2002kk}
  G.~W.~Semenoff and K.~Zarembo,
  Nucl.\ Phys.\ Proc.\ Suppl.\  {\bf 108}, 106 (2002)
  [arXiv:hep-th/0202156].

\bibitem{Lin:2006rf}
  S.~Lin and E.~Shuryak,
  Phys.\ Rev.\  D {\bf 77}, 085013 (2008)
  [arXiv:hep-ph/0610168].

\bibitem{Hofman:2008ar}
  D.~M.~Hofman and J.~Maldacena,
  JHEP {\bf 0805}, 012 (2008)
  [arXiv:0803.1467 [hep-th]].

\bibitem{Peeters:2005fq}
  K.~Peeters, J.~Sonnenschein and M.~Zamaklar,
  JHEP {\bf 0602}, 009 (2006)
  [arXiv:hep-th/0511044].


\bibitem{Shuryak:2005ia}
  E.~Shuryak, S.~J.~Sin and I.~Zahed,
  J.\ Korean Phys.\ Soc.\  {\bf 50}, 384 (2007)
  [arXiv:hep-th/0511199].

\bibitem{Lin:2008rw}
  S.~Lin and E.~Shuryak,
  Phys.\ Rev.\  D {\bf 78}, 125018 (2008)
  [arXiv:0808.0910 [hep-th]].

\bibitem{Gubser:2008pc}
  S.~S.~Gubser, S.~S.~Pufu and A.~Yarom,
  Phys.\ Rev.\  D {\bf 78}, 066014 (2008)
  [arXiv:0805.1551 [hep-th]].

\bibitem{Lin:2009pn}
  S.~Lin and E.~Shuryak,
  Phys.\ Rev.\  D {\bf 79}, 124015 (2009)
  [arXiv:0902.1508 [hep-th]].

\bibitem{Gubser:2009sx}
  S.~S.~Gubser, S.~S.~Pufu and A.~Yarom,
  JHEP {\bf 0911}, 050 (2009)
  [arXiv:0902.4062 [hep-th]].

\bibitem{Lin:2010cb}
  S.~Lin and E.~Shuryak,
  Phys.\ Rev.\  D {\bf 83}, 045025 (2011)
  [arXiv:1011.1918 [hep-th]].


\bibitem{Lublinsky:2007mm}
  M.~Lublinsky, E.~Shuryak,
  Phys.\ Rev.\  {\bf C76}, 021901 (2007).
  [arXiv:0704.1647 [hep-ph]]

\bibitem{Chesler:2010bi}
  P.~M.~Chesler and L.~G.~Yaffe,
  Phys.\ Rev.\ Lett.\  {\bf 106}, 021601 (2011)
  [arXiv:1011.3562 [hep-th]].

\bibitem{janik2}
  M.~P.~Heller, R.~A.~Janik and P.~Witaszczyk,
  ``The characteristics of thermalization of boost-invariant plasma from
  holography,''
  arXiv:1103.3452 [hep-th].

\bibitem{Aamodt:2010pb}
  B.~Abelev {\it et al.}  [The ALICE Collaboration],
  Phys.\ Rev.\ Lett.\  {\bf 105}, 252301 (2010)
  [arXiv:1011.3916 [nucl-ex]].


\bibitem{Lublinsky:2011cw}
  M.~Lublinsky, E.~Shuryak,
  [arXiv:1108.3972 [hep-ph]].

\bibitem{Mino:1996nk}
  Y.~Mino, M.~Sasaki, T.~Tanaka,
  Phys.\ Rev.\  {\bf D55}, 3457-3476 (1997).
  [gr-qc/9606018].
  
\bibitem{Quinn:1996am}
  T.~C.~Quinn, R.~M.~Wald,
  Phys.\ Rev.\  {\bf D56}, 3381-3394 (1997).
  [gr-qc/9610053]


\bibitem{Shuryak:2011ge}
  E.~Shuryak, H.~U.~Yee and I.~Zahed,
  arXiv:1110.0825 [hep-th].

\bibitem{Khriplovich:1973bq}
  I.~B.~Khriplovich and E.~V.~Shuryak,
  Zh.\ Eksp.\ Teor.\ Fiz.\  {\bf 65}, 2137 (1973).

\bibitem{Shuryak:2002ai}
  E.~V.~Shuryak and I.~Zahed,
  Phys.\ Rev.\  D {\bf 67}, 054025 (2003)
  [arXiv:hep-ph/0207163].


\end{thebibliography}
\end{document}